\documentclass[12pt]{article}
\usepackage{amssymb,amsmath,latexsym,graphicx, bm, mathrsfs, hyperref}

\begin{document}

\vspace{-0.5cm}
\title{\vspace{-1.5cm} Geometrical Formulation of Relativistic Mechanics \vspace{-0.25cm}\author{ Sumanto Chanda$^1$, Partha Guha$^{1,2}$}}

\maketitle
\thispagestyle{empty}

\vspace{-0.35cm}
\begin{minipage}{0.46\textwidth}
\begin{flushleft}
\textit{\small $ ^1$ S.N. Bose National Centre for Basic Sciences} \\
\textit{\small JD Block, Sector-3, Salt Lake, \\ Kolkata-700098, INDIA.} 
\end{flushleft}
\end{minipage}
\begin{minipage}{0.46\textwidth}
\begin{flushright}
\textit{\small $^2$ Instituto de F\'isica de S\~ao Carlos; IFSC/USP} \\
\textit{\small Universidade de S\~ao Paulo Caixa Postal 369,} \\
\textit{\small CEP 13560-970, S\~ao Carlos-SP, Brazil}
\end{flushright}
\end{minipage} 

\begin{center}
\texttt{\small sumanto12@bose.res.in ,  partha@bose.res.in}
\end{center}


\smallskip


formulation, relativistic Kepler, Bohlin-Arnold duality, relativistic Chiellini condition, Lienard equation.

\abstract{The relativistic Lagrangian in presence of potentials was formulated directly from the 
metric, with the classical Lagrangian shown embedded within it. Using it we formulated 
covariant equations of motion, a deformed Euler-Lagrange equation, and relativistic 
Hamiltonian mechanics. We also formulate a modified local Lorentz transformation, 
such that the metric at a point is invariant only under the transformation defined at that 
point, and derive the formulae for time-dilation, length contraction, and gravitational 
redshift. Then we compare our formulation under non-relativistic approximations to the 
conventional ad-hoc formulation, and we briefly analyze the relativistic Li\'enard oscillator 
and the spacetime it implies.}

\vspace{-0.5cm}
\tableofcontents
\setlength{\topmargin}{-.70in}  % Top margin of 2 in -0.75 in = 1 in
\setlength{\textheight}{8.75in}  % Lower margin of 11 in - 9 in - 1 in = 1 in

\newpage
\setcounter{page}{1}

\numberwithin{equation}{section}

\section{Introduction}	

In relativistic mechanics, we describe a geometric perspective of dynamics. This means that 
we start by describing pseudo-Riemannian spaces via metrics $(M, g)$ by which we shall measure 
infinitesimal arc lengths in such spaces. Dynamical trajectories or geodesics between any two 
chosen fixed points are the shortest path in terms of integrated length in between. We are 
familiar with the usage of infinitesimal arc length in special relativity for flat spaces.
\begin{equation}
\label{flatmet} ds^2 = \eta_{\mu \nu} \ dx^\mu dx^\nu = c^2 dt^2 - | d {\bm x} |^2.
\end{equation}
In general or curved spaces, the general infinitesimal arc length element is given by:
\begin{equation}
\label{genmet} ds^2 = g_{\mu \nu} dx^\mu dx^\nu = g_{00} c^2 dt^2 + 2 g_{0i} c \; dt \; dx^i + g_{ij} dx^i dx^j, \qquad i, j = 1, 2, 3.
\end{equation}
which becomes flat when $g_{\mu \nu} = \eta_{\mu \nu} = \text{diag}(1, -1, -1, -1)$. Thus, we must derive mechanical 
formulation from (\ref{genmet}) to correctly describe relativistic mechanics in general on curved spaces. 
Spacetime for the usual problems dealt with in classical mechanics simply involve a 4-potential 
$A^\mu = (U, - {\bm A})$, for which, we shall have the spatial terms of the metric are flat ($g_{ij} = - \delta_{ij}$). \smallskip

One important rule for a metric to abide by is that it must be invariant under 
the Lorentz transformation, which is easily formulated in special relativity for free 
particles travelling at constant velocity. However, the familiar Lorentz transformation 
does not preserve metrics describing trajectories for particles in the presence 
of potential fields. Thus we must define the Lorentz transformation in such a way 
that it locally preserves such metrics, meaning that the metric at a point is invariant 
only under the transformation rule defined at that same point. Its local nature 
means that position co-ordinates cannot be transformed as in special relativity. \smallskip

There is an alternate ad hoc approach taken to formulate the relativistic 
Lagrangian, employed in publications by Harvey \cite{harvey} and Babusci et al \cite{babusci}, to describe the 
dynamics of the relativistic oscillator and Kepler. It essentially draws from the conventional 
formulation of the Lagrangian provided by Goldstein in \cite{goldstein}
\begin{equation}
\label{adhoc} L = - mc^2 \sqrt{1 - \left( \frac vc \right)^2} - U.
\end{equation}
where the kinetic energy term is kept seperate from the potential term. Although (\ref{adhoc}) gives 
the correct relativistic answers for practical problems, this Lagrangian is not Lorentz covariant.
One may suspect that under some approximations, the formulation born from the metric will 
transform into the conventional Lagrangian based approach. 

The conventional gravitational field yields conserved dynamics in a central force field without 
drag. However, motion involving drag is a significant topic in the study of dynamical systems, 
describing realistic situations, with practical applications mostly in engineering. It 
would naturally be very interesting to see what kinds of spacetimes and gravitational fields 
produce dynamics involving drag. Such systems are not always integrable unless certain 
conditions are fulfilled by the drag co-efficient functions and the force-field functions. The relativistic 
generalizations of Lagrangian/Hamiltonian systems with position-dependent mass \cite{bgsn, behr1, behr2} 
could be treated within this formalism. Such systems are in some cases equivalent to constant 
mass motion on curved spaces, and some nonlinear oscillators can be interpreted in this setting. 

\smallskip

This paper is organized as follows: Section 2 is devoted to the preliminaries on the 
formulation of special relativistic mechanics in flat spacetime, followed by static 
curved spaces dealt with in classical mechanics. There we deduce the relativistic 
deformation of the Euler-Lagrange equation, and a conserved quantity related to such 
mechanics. We also describe relativistic Hamiltonian mechanics in curved spaces.

Section 3 deals with the modification of the Lorentz transformation under which such 
metrics are invariant. This is necessary since the regular Lorentz transformation, 
designed to work for free particles in the case of special relativity, will not suffice for 
particles accelerating under the influence of a potential field. Then, we will deduce 
the formulas for time-dilation, length contraction, and gravitational redshift from 
the modified Lorentz transformation formula.

Section 4 will list the various approximations that can be made and how they affect 
our formulations. Here we will verify if the relativistic Lagrangian under any of these approximations 
transforms into the conventional one employed by Harvey and Babusci et al. 

Section 5 will cover formulation of the relativistic 2D oscillator using our approach. Here 
we will verify if the Bohlin-Arnold-Vasiliev duality between relativistic Kepler and Hooke systems 
holds in such non-classical settings, and what approximations, if any, are required. Such 
dualities are observable in Bertrand Spacetime metrics which correspond either to oscillator 
or to Kepler systems on the associated three dimensional curved spatial manifold \cite{behr3}, which 
could be studied from this persepective.

Section 6, finally, will study relativistic damped mechanical systems and redefine the Chiellini 
integrability condition in relativistic form. Then we will deduce the related metric for damped 
systems and define its contact Hamiltonian structure.

\section{Preliminary: Relativistic Mechanics}

Here, we shall describe relativistic formulation first for free particle, then for classical static 
curved spaces with scalar potentials. The general spacetime metric for curved spaces in the 
locality of potential sources can be formulated by defining (\ref{genmet}) as a perturbation around the 
flat spacetime metric (\ref{flatmet}) that asymptotically vanishes towards infinity:
\begin{equation}
\label{pert} g_{\mu \nu} (x) = \eta_{\mu \nu} + h_{\mu \nu} (x) \qquad \qquad \lim_{x \rightarrow \infty} h_{\mu \nu} (x) = 0
\end{equation}
This is necessary to ensure that the metric is asymptotically flat at large distances from the 
potential field source.
$$\lim_{x \rightarrow \infty} ds^2 = \lim_{x \rightarrow \infty} g_{\mu \nu} (x) \ dx^\mu dx^\nu = \eta_{\mu \nu} dx^\mu dx^\nu$$
After that, we will discuss the Hamiltonian mechanical description. Then we shall explore 
various approximations and the modified results that follow.

\numberwithin{equation}{subsection}

\subsection{Flat space}

In special relativity, we are discussing free particle mechanics $U ({\bm x}) = 0$. This means that the 
metric describes flat space (\ref{flatmet}):
$$ds^2 = \eta_{\mu \nu} dx^\mu dx^\nu = c^2 dt^2 - \big| d \bm{x} \big|^2$$
from which the Lagrangian derived is:
$$\widetilde{\mathcal{L}} = - mc \sqrt{\left( \frac{ds}{d \tau} \right)^2} = - mc \sqrt{c^2 \dot{t}^2 - \big| \dot{\bm{x}} \big|^2} = - mc^2 \dot{t} \sqrt{1 - \left( \frac{| \bm{v} |}c \right)^2} \qquad \text{where } \quad \bm{v} = \dfrac{\dot{\bm x}}{\dot{t}}.$$ 
Alternatively, we can define the reparameterised Lagrangian $\mathcal{L}$ from the geometric action as:
$$S = \int_1^2 d \tau \widetilde{\mathcal{L}} = - mc^2 \int_1^2 dt \sqrt{1 - \beta^2} = \int_1^2 dt \ \mathcal{L}, \qquad \qquad \text{where } \beta = \frac{|\bm v|}c,$$
$$\Rightarrow \qquad \mathcal{L} = - mc \sqrt{\left( \frac{ds}{d t} \right)^2} = - mc^2 \sqrt{1 - \beta^2}.$$
This results in the relativistic momentum and energy given as:
\[ p_\mu = \dfrac{\partial \widetilde{\mathcal{L}}}{\partial \dot x^\mu} = 
\begin{cases}
\bm{p} = \ \ \dfrac{\partial \widetilde{\mathcal{L}}}{\partial \dot{\bm x}} = \dfrac{m {\bm v}}{\sqrt{1 - \beta^2}} = m {\bm v} \gamma \\ 
\mathcal{E} = - \dfrac{\partial \widetilde{\mathcal{L}}}{\partial \dot{t}} = \dfrac{m c^2}{\sqrt{1 - \beta^2}} = m c^2 \gamma = \dfrac{\bm{p} . \dot{\bm x} - \widetilde{\mathcal{L}}}{\dot t} = \bm{p} . {\bm v} - \mathcal{L}
\end{cases} \]
where, we have designated a reparameterising factor:
\begin{equation}
\label{nrgamma} \gamma = \frac1{\sqrt{1 - \beta^2}}.
\end{equation}
from which, we have the familiar equations for relativistic energy:
\begin{equation}
\label{flaten} \mathcal{E}^2 = \eta^{\mu \nu} p_\mu p_\nu = |p|^2 c^2 + m^2 c^4.
\end{equation}
Furthermore, the singularity that occurs when $\left| \bm v \right| = c$ in the denominator in (\ref{nrgamma}), ensures 
that the speed of light is never exceeded, establishing is as the physical upper limit of velocity 
in flat spaces as elaborated in \cite{dr1, dr2}. This concludes the basics of flat space. From here on, 
we will use the alternative convention involving mechanical systems parameterized wrt time. 
Next we shall look at curved spaces with scalar potentials.

\subsection{Curved space for single scalar potential}

When considering mechanics on classical static curved spaces, where the source of spacetime 
curvature is one scalar potential source $U ({\bm x}) \neq 0$, the metric based arc length can also be used 
for curved spaces by including the potential $U ({\bm x})$ into the metric as a specific version of (\ref{genmet}) 
following Gibbons' prescription \cite{gwg, cgg} as shown below:
\begin{equation}
\label{curmet} ds^2 = g_{00} (\bm{x}) c^2 dt^2 - \big| d \bm{x} \big|^2 \qquad \qquad \text{where} \quad g_{00} (\bm{x}) = 1 + \frac{2 U ({\bm x})}{mc^2}.
\end{equation}
where the potential $U ({\bm x})$ is a factor in the only non-zero perturbative term $h_{00} (x)$ of the 
temporal metric term $g_{00} ({\bm x})$ that vanishes asymptotically, while $h_{ij} (\bm x) = 0 \ \forall \ i, j = 1, 2, 3$. 
The relativistic action $S$ and Lagranigan $\mathcal{L}$ are:
$$S = \int_1^2 d \tau \ {\mathcal L} \qquad \qquad \mathcal{L} = - mc \sqrt{\left( \frac{ds}{d \tau} \right)^2} = - mc \sqrt{g_{00} (\bm{x}) c^2 \dot t^2 - \left| \bm{\dot x} \right|^2}.$$
If we include the potential linearly as a perturbation into the metric, then we have
$$\mathcal{L} = - mc^2 \sqrt{\left( 1 + \frac{2 U (\bm{x})}{mc^2} \right) \dot t^2 - \left( \frac{| \bm{\dot x} |}c \right)^2} = - mc^2 \dot t \sqrt{1 - \frac 2{mc^2} \left( \frac{m | {\bm v} |^2}2 - U (\bm{x}) \right)}.$$
Now, we are familiar with the traditional non-relativistic or classical Lagrangian
\begin{equation}
\label{classlag} L = T - U = \frac{m | {\bm v} |^2}2 - U (\bm{x}), \qquad \qquad T = \frac{m | {\bm v} |^2}2.
\end{equation}
Thus, the relativistic Lagrangian $\mathcal{L}$ is given by:
\begin{equation}
\label{rlag} \mathcal{L} = - mc \sqrt{\left( \frac{ds}{d \tau} \right)^2} = - mc^2 \sqrt{1 - \frac {2 L}{mc^2}}.
\end{equation}
Under the circumstances that we are dealing with a stationary free particle, we can define the 
ground-state relativistic Lagrangian as follows:
\[ \begin{split}
{\bm v} = 0 \quad \\ 
\quad U({\bm x}) = 0
\end{split}
 \Bigg\} \quad \Rightarrow \quad L = 0 \quad \Rightarrow \quad {\mathcal L} = L_0 = - mc^2. \]
Thus, we can re-write the relativistic Lagrangian as follows:
\begin{equation}
\label{curlag} \mathcal{L} = L_0 \sqrt{1 + 2 \frac L{L_0}}.
\end{equation}
showing that the classical Lagrangian $L$ is embedded within the relativistic Lagrangian $\mathcal{L}$. 
Furthermore, we can say that we recover flat space (\ref{flaten}) when $g_{00} (\bm{x}) =1$ ie. $(U (\bm{x}) = 0)$.
\begin{equation}
\label{repara} \frac{d s}{d t} = c \sqrt{1 + 2 \frac {L}{L_0}} \quad \Rightarrow \quad \Gamma^{-1} = \frac{d \widetilde{t}}{d t} = \sqrt{1 + 2 \frac {L}{L_0}} \quad \xrightarrow{U = 0} \quad \gamma^{-1} = \sqrt{1 - \left( \frac{|\bm v|}c \right)^2}.
\end{equation}
while the relativistic momenta are clearly Lorentz-covariant, with $\widetilde{t}$ being the proper time in 
the particle frame. 
\begin{equation}
\label{relmom} p_\mu = \frac{\partial \mathcal{L}}{\partial \dot x^\mu} = 
\begin{cases}
\bm{p} = \ \ \dfrac{\partial \mathcal{L}}{\partial \dot{\bm x}} = m {\bm v} \Gamma \smallskip \\
\mathcal{E} = - \dfrac{\partial \mathcal{L}}{\partial \dot{t}} = m c^2 g_{00} (\bm{x}) \Gamma = \dfrac{\bm{p} . \dot{\bm x} - \widetilde{\mathcal{L}}}{\dot t}
\end{cases}
\end{equation} 
From the momenta (\ref{relmom}), we can see that the singularity that occurs when $\Gamma^{-1} = 0$ establishes a different physical upper limit for the velocity of a particle:
$$1 + 2 \frac L{L_0} = 0 \qquad \Rightarrow \qquad g_{00} (\bm x) c^2 {\dot t}^2 - \left| \dot{\bm x} \right|^2 = 0 \qquad \Rightarrow \qquad \left| \bm v_{\null} \right| = \left| \frac{d \bm x}{d t} \right| = c \sqrt{g_{00} (\bm x)}.$$
which is less than the speed of light $c$ since $g_{00} (\bm x) < 1$, ($U (\bm x) \leq 0$ for gravitational 
potentials). This deformation of the relativistic 4-momentum as a consequence of 
the change of the speed limit, is due to the change in length contraction and time-
dilation due to the gravitational potential under a local Lorentz transformation. \\ 

\noindent
Under the approximation $L << L_0$, binomially expanding (\ref{curlag}) :
\begin{equation}
\label{nrlag} \mathcal{L} \xrightarrow{L << mc^2 = - L_0} L_0 \left( 1 + \frac L{L_0} \right) = L + L_0.
\end{equation}
Thus, in the non-relativistic limit, $L$ given by (\ref{classlag}) will suffice to produce the equations of 
motion. One can alternatively say that the effective classical Lagrangian directly derived from 
the metric (\ref{curmet}) is
$$\mathcal{L} \xrightarrow{L << mc^2 = - L_0} - mc^2 \left( 1 - \frac L{mc^2} \right) = L - mc^2 = - \frac m2 \left( \frac{ds}{dt} \right)^2 - \frac{mc^2}2,$$
\begin{equation}
\label{efflag} L_{eff} = - \frac m2 \left( \frac{ds}{dt} \right)^2 = \frac m2 \left( \big| \dot{\bm x} \big|^2 - c^2 g_{00} (\bm{x}) \right) \qquad \qquad g_{00} (\bm{x}) = 1 + \frac{2 U (\bm{x})}{mc^2}.
\end{equation}
We shall now proceed to analyze the relativistic equations of motion.

\subsubsection{Relativistic equations of motion}

Now we shall turn our attention to formulating of the equations of motion. The Euler-Lagrange 
equation is given by:
$$\frac{d \ }{d \tau} \left( \frac{\partial \mathcal{L}}{\partial \dot{x}^i} \right) = \frac{\partial \mathcal{L}}{\partial x^i}.$$
when applied to (\ref{curlag}), we get the relativistic equation of motion:
\begin{equation}
\label{intermed1} \frac{d \ }{d t} \left( m \Gamma {\bm v} \right) = - \Gamma \bm{\nabla} U (\bm{x}).\frac{d \ }{d t} \left( m \Gamma {\bm v} \right) = - \Gamma \bm{\nabla} U (\bm{x}).
\end{equation}
then we can write using (\ref{repara}): \quad $\widetilde{\bm v} = \dfrac{d \bm{x}}{d \widetilde{t}} = \dfrac{d t}{d \widetilde{t}} {\bm v} = \Gamma {\bm v}$, making (\ref{intermed1}) into
\begin{equation}
\label{eom} m \frac{d^2 \bm{x}}{d {\widetilde t}^2} = \Gamma \frac{d \ }{d t} \left( m \Gamma {\bm v} \right) = - \Gamma^2 \bm{\nabla} U (\bm{x}).
\end{equation}
If we expand $\Gamma^2$ in this equation, we will get: 
$$\Gamma^2 = \left( 1 + 2 \frac L{L_0} \right)^{-1} = 1 - 2 \frac L{L_0} + ( - 2 )^2 \left( \frac L{L_0} \right)^2 + . . . . $$, 
$$\therefore \qquad m \frac{d^2 \bm{x}}{d {\widetilde t}^2} = - \bm{\nabla} U (\bm{x}) + 2 \frac L{L_0} \bm{\nabla} U (\bm{x}) - . . . . . $$
which is the equation of motion with additional terms associated with the force function. 
Alternatively, we can say that on applying the Euler-Lagrange equation to the relativistic 
Lagrangian (\ref{curlag}), we will get:
$${\mathcal L} = L_0 \sqrt{1 + 2 \frac L{L_0}}, \qquad \qquad \frac{\partial \mathcal L}{\partial v^i} = \frac{L_0}{2 \mathcal L} \frac{\partial L}{\partial v^i} \qquad \qquad \frac{\partial \mathcal L}{\partial x^i} = \frac{L_0}{2 \mathcal L} \frac{\partial L}{\partial x^i},$$
$$ \frac{d \ }{d t} \left( \frac{\partial \mathcal L}{\partial v^i} \right) = \frac{L_0}{2 \mathcal L} \frac{d \ }{d t} \left( \frac{\partial L}{\partial v^i} \right) - \left( \frac{\partial L}{\partial v^i} \right) \frac{L_0^3}{2 \mathcal L^3} \frac{d L}{d t} .$$ 
Recalling (\ref{repara}), and writing $\Gamma = \dfrac{L_0}{\mathcal L}$ according to (\ref{curlag}), we get from the Euler-
Lagrange equation $\frac{d \ }{d t} \left( \frac{\partial \mathcal L}{\partial v^i} \right) - \frac{\partial \mathcal{L}}{\partial x^i} = 0$
$$\frac{L_0}{2 \mathcal L} \left[ \frac{d \ }{d t} \left( \frac{\partial L}{\partial v^i} \right) - \frac{\partial L}{\partial x^i} \right] - \left( \frac{\partial L}{\partial v^i} \right) \frac{L_0^3}{2 \mathcal L^3} \frac{d L}{d t} = 0,$$
\begin{equation}
\label{neweuler} \therefore \qquad \left[ \frac{d \ }{d t} \left( \frac{\partial L}{\partial v^i} \right) - \frac{\partial L}{\partial x^i} \right] = \Gamma^2 \left( \frac{\partial L}{\partial v^i} \right) \frac{d L}{d t} \qquad \qquad \Gamma = \frac{L_0}{\mathcal L}.
\end{equation}
Thus, we have a relativistic deformation (\ref{neweuler}) of the Euler-Lagrange equation for the classical 
Lagrangian $L$ derived by applying the original Euler-Lagrange equation to the relativistic 
Lagrangian $\mathcal L$. 

\smallskip

Under non-relativistic limits, $\Gamma \approx 1$, (\ref{eom}) becomes the more familiar form of the 
equation of motion given below that directly derive from Euler-Lagrange equations applied 
upon (\ref{classlag}):
\begin{equation}
\label{classeom} m \frac{d^2 \bm{x}}{d t^2} = - \bm{\nabla} U (\bm{x}).
\end{equation}
Similarly, the Euler-Lagrange equation that derives from the effective Lagrangian (\ref{efflag}) is:
\begin{equation}
\label{effeom} \ddot{\bm x} = - \frac{c^2}2 {\bm \nabla} g_{00} (\bm{x}) \equiv - \frac1m {\bm \nabla} U (\bm{x}).
\end{equation}
which is equivalent to (\ref{classeom}). Under the circumstances that one has either the 
relativistic or classical equations of motion, there is an algorithm to deduce the 
spacetime metric that generates such dynamics:

\begin{enumerate}
\item Convert the relativistic equation to the non-relativistic version (\ref{classeom}) (with $\gamma = 1$).
\item Deduce $U (\bm{x})$ from the non-relativistic equation (\ref{classeom}).
\item Use $U$ to reproduce the spacetime metric according to (\ref{efflag}).
\end{enumerate}
This algorithm allows us to formulate a curved spacetime metric that reproduces the dynamics 
described by the equation of motion considered. If the force is not gravitational in nature, it 
essentially produces the equivalent curved spacetime that can act as a gravitational substitute 
for the mechanical system that imitates its observed motion classically. 

\subsubsection{Conserved Quantity}

We shall now regard a conserved quantity deduced from (\ref{eom}) comparable to the Hamiltonian and the geodesic 
equation of motion. The conserved quantity in question is:
\begin{equation}
\label{cnsv} \therefore \qquad K^2 = \frac{\left( 1 + \dfrac{2 U}{mc^2} \right)^2}{1 - \dfrac {2 L}{mc^2}} = \left( \Gamma g_{00} ({\bm x}) \right)^2.
\end{equation}
where $K$ is essentially a multiple of the relativistic energy given by (\ref{relmom}). Applying (\ref{cnsv}) 
to (\ref{repara}), we can say that
$$\Gamma^2 = \left( \frac{d t}{d \widetilde{t}} \right)^2 = \frac1{1 - \frac {2 L}{mc^2}} = \frac{K^2}{\left( 1 + \frac{2 U}{mc^2} \right)^2}.$$
Along a geodesic, the equation of motion in the particle frame (\ref{eom}) can be re-written as
$$m \frac{d^2 \bm{x}}{d\widetilde{t}^2} = - \left( \frac{d t}{d \widetilde{t}} \right)^2 \bm{\nabla} U (\bm{x}) = - \frac{mc^2}2 \frac{K^2}{\left( 1 + \frac{2 U}{mc^2} \right)^2} \bm{\nabla} \left( 1 + \frac{2U}{mc^2} \right) = \frac{mc^2}2 K^2 \bm{\nabla} \left( 1 + \dfrac{2U}{mc^2} \right)^{-1},$$
\begin{equation}
\label{geoeom} \therefore \qquad \frac{d^2 \bm{x}}{d\widetilde{t}^2} = \frac{K^2 c^2}2 \bm{\nabla} \left( g_{00} ({\bm x}) \right)^{-1} = \frac{c^2}2 \bm{\nabla} \left( \Gamma^2 g_{00} ({\bm x}) \right).
\end{equation}
which is one way of writing the relativistic equation of motion. If we are consider Lorentz transformations of space-time event intervals, and $g_{00} = g_{00} (\bm{x})$, we 
can say that
$$dt \longrightarrow d \widetilde{t} = \frac{d \widetilde{t}}{d t} dt \qquad \Rightarrow \qquad g_{00} \longrightarrow \widetilde{g}_{00} = \left( \frac{d \widetilde{t}}{d t} \right)^{-2} g_{00} = \Gamma^2 g_{00} (\bm{x}).$$
which lets us write the Lorentz-covariant equation of motion:
\begin{equation}
\label{coveom} \frac{d^2 \bm{x}}{d\widetilde{t}^2} = \frac{c^2}2 \bm{\nabla} \widetilde{g_{00}} (\bm{x}).
\end{equation}
Now we shall reproduce the known and familiar relativistic phenomena of time dilation and 
gravitational red-shift from this formulation as a test.

\subsection{Relativistic Hamiltonian mechanics in curved spaces}

The Hamiltonian formulation of classical mechanics is very useful, not just for its geometrical 
properties, but also for enabling extension of the classical theory into the quantum 
context via standard quantization. Having described a relativistic Lagrangian formulation for 
mechanics in the presence of a scalar potential, it is 
natural to also consider the Hamiltonian formulation. \\ \\ 
Referring to (\ref{relmom}), the relativistic energy for curved spaces is:
\begin{equation}
\label{relen1} \mathcal{E} = \left[ m c^2 + 2 U (\bm{x}) \right] \Gamma \qquad \qquad \mathcal{E}^2 = g_{00} (\bm{x}) \left( |{\bm p}|^2 c^2 + m^2 c^4 \right).
\end{equation}
This would effectively make the Hamiltonian $\mathcal{H}$:
$$\mathcal{H} = \sqrt{g_{00} (\bm{x})} \sqrt{|p|^2 c^2 + m^2 c^4}.$$
and the Hamilton's equations of motion for $g_{00} (\bm{x}) = 1 + \dfrac{2 U (\bm{x})}{mc^2}$:
\begin{equation} \label{hameq1}
\begin{split}
\dot{\bm x} &= \frac{\partial \mathcal{H}}{\partial \bm{p}} = \sqrt{g_{00} (\bm{x})} \frac{c \bm{p}}{\sqrt{|p|^2 + m^2 c^2}}, \\
\dot{\bm p} &= - \frac{\partial \mathcal{H}}{\partial \bm{x}} = - \frac{\bm{\nabla} U (\bm{x})}{\sqrt{g_{00} (\bm{x})}} \sqrt{ \left( \frac{|p|}{mc} \right)^2 + 1}.
\end{split}
\end{equation}
Such formulation has been applied to study the relativistic version of the Quantum Harmonic 
Oscillator \cite{poszwa}. In the following section, we will elaborate on the Lorentz transformation 
operation for such metrics.

\numberwithin{equation}{section}

\section{A modified Local Lorentz Transformation}

An important issue that emerges is the invariance of such metrics under Lorentz transformations. The Lorentz transformation we are familiar with applies only to special relativity, where we deal with 
free particles.  \\ \\
The conventional Lorentz boost of co-ordinates $x$ of frame $F$ to $\widetilde x$ of frame $\widetilde F$ is:
\begin{equation} \label{srlorentz}
\begin{split}
c \widetilde t &= \gamma c t - \gamma \beta x \\
\widetilde x &= \gamma \beta c t - \gamma x
\end{split} \qquad \qquad \qquad 
\begin{split}
x &= vt, \\ 
\gamma = \frac1{\sqrt{1 - \beta^2}}, &\qquad \beta = \frac{V_{\widetilde F F}}c
\end{split}.
\end{equation}
where $v$ is the constant velocity of the particle in frame $F$, and  $V_{\widetilde F F}$ is the speed of frame $\widetilde F$ with respect to frame $F$. 
A better way to write (\ref{srlorentz}) this locally is to replace: $x^\mu \rightarrow d x^\mu$ under which the 
spacetime metric is invariant.
\begin{equation} \label{local} 
\begin{split}
c \ d \widetilde t &= \gamma \ c \ d t - \gamma \beta \ d x \\
d \widetilde x &= \gamma \beta \ c \ dt - \gamma \ d x
\end{split},
\end{equation}
$$d \widetilde s^2 = c^2 d \widetilde t^2 - d \widetilde x^2 \quad = \quad ds^2 = c^2 d t^2 - d x^2.$$
The scenario we are dealing with in this case involves a particle under the influence of a 
potential field. The metric is easily invariant under rotations in the presence of spherically 
symmetric potentials, which leaves only boosts to be considered. Due to the presence of a 
potential, we are required to use a modified Lorentz boost operation, which we shall briefly 
derive here. The Lorentz boost equations are:
\begin{equation} \label{modlor} 
\begin{split}
c \ d \widetilde t &= \Lambda^0_0 \ c \ d t + \Lambda^0_1 \ d x \\
d \widetilde x &= \Lambda^1_0 \ c \ dt + \Lambda^1_1 \ d x
\end{split} 
\qquad \qquad \Lambda = \left( {\begin{array}{cc} \Lambda^0_0 & \Lambda^0_1 \\ \Lambda^1_0 & \Lambda^1_1 \end{array}} \right).
\end{equation}
If we consider a Lorentz transformation to the particle frame, we should have $d \widetilde x = 0$, which 
means that from the second equation of (\ref{modlor}), we have:
\begin{equation}
\label{rule1} \frac{\Lambda^1_0}{\Lambda^1_1} = \frac vc = \beta.
\end{equation}
Furthermore, the determinant of the matrix $\Lambda$ must be unity to preserve volume elements 
spanned by 4-vectors.
\begin{equation}
\label{rule2} \Lambda^0_0 \Lambda^1_1 - \Lambda^0_1 \Lambda^1_0 = 1.
\end{equation}
Now, demanding that the metric be invariant under the transformation gives us another rule:
$$\Lambda^t G \Lambda = G \qquad \Rightarrow \qquad \Lambda^t = G \Lambda^{-1} G^{-1}, \qquad G = \left( {\begin{array}{cc} g_{00} (\bm x) & 0 \\ 0 & - 1 \end{array}} \right), \qquad g_{00} (\bm x) = 1 + \frac{2 U (\bm x)}{mc^2},$$

\[ \begin{split} 
\Rightarrow \qquad \left( {\begin{array}{cc} \Lambda^0_0 & \Lambda^1_0 \\ \Lambda^0_1 & \Lambda^1_1 \end{array}} \right) &= \left( {\begin{array}{cc} g_{00} (\bm x) & 0 \\ 0 & - 1 \end{array}} \right) \left( {\begin{array}{cc} \Lambda^1_1 & - \Lambda^0_1 \\ - \Lambda^1_0 & \Lambda^0_0 \end{array}} \right) \left( {\begin{array}{cc} \left( g_{00} (\bm x) \right)^{-1} & 0 \\ 0 & - 1 \end{array}} \right), \\
&= \left( {\begin{array}{cc} \Lambda^1_1 & g_{00} (\bm x) \Lambda^0_1 \\ \left( g_{00} (\bm x) \right)^{-1} \Lambda^1_0 & \Lambda^0_0 \end{array}} \right),
\end{split} \]
\begin{equation}
\label{rule3} \Rightarrow \qquad  \Lambda^0_0 = \Lambda^1_1, \qquad \Lambda^1_0 = g_{00} (\bm x) \Lambda^0_1.
\end{equation}
Combining the equations (\ref{rule1}) and (\ref{rule3}) into (\ref{rule2}) gives us:
$$\left( \Lambda^1_1 \right)^2 - \left( g_{00} (\bm x) \right)^{-1} \left( \Lambda^1_0 \right)^2 = \left( \Lambda^1_1 \right)^2 \left( 1 - \left( g_{00} (\bm x) \right)^{-1} \beta^2 \right) = 1,$$
\begin{equation} \label{rule4} 
\begin{split}
\Lambda^0_0 &= \Lambda^1_1 = \frac1{\sqrt{1 - \left( g_{00} (\bm x) \right)^{-1} \beta^2}} = \sqrt{g_{00} (\bm x)} \Gamma, \\
\Lambda^1_0 &= g_{00} (\bm x) \Lambda^0_1 = \sqrt{g_{00} (\bm x)} \beta \Gamma.
\end{split}
\end{equation}
Thus, using (\ref{rule4}), the modified Lorentz boost matrix (\ref{modlor}) is given by: 
\begin{equation}
\label{1+1} \Lambda = \left( {\begin{array}{cc} \Gamma \sqrt{g_{00}} & - \beta \Gamma \left( \sqrt{g_{00}} \right)^{-1} \\ 
- \beta \Gamma \sqrt{g_{00}} & \Gamma \sqrt{g_{00}} \end{array}} \right),
\end{equation}

$$d \widetilde s^2 = \left( 1 + \frac{2 U (\bm x)}{mc^2} \right) c^2 d \widetilde t^2 - d \widetilde x^2 \quad = \quad ds^2 = \left( 1 + \frac{2 U (\bm x)}{mc^2} \right) c^2 d t^2 - d x^2.$$
Thus, we have a modified local Lorentz transformtion that preserves the metric. For a $3 + 1$ 
spacetime, the modified local Lorentz boost matrix between one co-ordinate and time would 
be written as: 

\begin{equation}
\label{3+1} \Lambda = \left( {\begin{array}{cccc} \Gamma \sqrt{g_{00}} & - \beta \Gamma \left( \sqrt{g_{00}} \right)^{-1} & 0 & 0 \\ - \beta \Gamma \sqrt{g_{00}} & \Gamma \sqrt{g_{00}} & 0 & 0 \\ 0 & 0 & 1 & 0 \\ 0 & 0 & 0 & 1 \end{array}} \right).
\end{equation}

The local nature of this transformation should not be surprising since according to the 
Equivalence principle, at any point on a curved manifold, there always exists a diffeomorphism 
that transforms it locally into a flat manifold. This means that alternatively, within the locality 
of that point in the local inertial frame, we can transform the problem into special relativity 
and apply the regular Lorentz transformation (\ref{local}). 

Naturally, it is not possible to perform a global co-ordinate transformation like (\ref{srlorentz}). In 
fact, we must understand that (\ref{srlorentz}) derives from (\ref{local}) via integration, and not the other 
way around via differentiation. This is because special relativity, as the name implies, describes 
a special case where spacetime is isotropic due to the absence of any potentials and any global 
co-ordinates are described as $x^\mu = \int_1^2 d \tau \ \dot{x}^\mu = \dot{x}^\mu \tau$. 

\numberwithin{equation}{subsection}
\subsection{Time-dilation and length contraction}

One of the reasons a speed limit exists is the Minkowsian signature of spacetime, 
which allows null geodesics for non-null spacetime intervals. This results in phenomena 
such as time-dilation and length contraction. The inclusion of a gravitational 
potential results in a modification of such phenomena, just as it has resulted in 
modification of the Lorentz transformation. \smallskip

Since we are discussing a modified local Lorentz transformation (\ref{3+1}), let us 
define the local co-ordinates in the neighbourhood of $x$ where $g_{00}$ is 
roughly constant:
\begin{equation}
\label{localcord} x \longrightarrow x + \chi, \qquad t \longrightarrow t + \tau.
\end{equation}
Now consider a problem where in a frame $S$, a stationary particle lies at a position $x = L$ along the $x$-axis, and $S$ moves at a velocity $v = \beta c$ along the $x$-axis wrt an observer in frame $\widetilde S$. Using (\ref{localcord}) the local Lorentz transformation equations from $S$ to $\widetilde S$ are:
\begin{equation} \label{lorp} 
\begin{split}
c \ d \widetilde \tau &= \ \ \Gamma \sqrt{g_{00} (x)} \ c \ d \tau + \beta \Gamma \left( \sqrt{g_{00} (x)} \right)^{-1} \ d \chi \\
d \widetilde \chi &= \beta \Gamma \sqrt{g_{00} (x)} \ c \ d \tau + \ \ \Gamma \sqrt{g_{00} (x)} \ d \chi
\end{split} 
\end{equation}
Since the particle is stationary in $S$ ($d \chi = 0$), the first equation of (\ref{lorp}) gives us:
\begin{equation}
\label{tdilp} d \widetilde \tau = \Gamma \sqrt{g_{00} (x)} d \tau \qquad \Rightarrow \qquad \boxed{ d \widetilde \tau = \Gamma \sqrt{g_{00} (x)} d \tau}.
\end{equation}
Integration of (\ref{lorp}) including constants of integration yields:
\begin{equation} \label{lorintp}
\begin{split}
c \widetilde \tau &= \ \ \Gamma \sqrt{g_{00} (x)} \ c \left( \int d \tau + \tau_0 \right) + \beta \Gamma \left( \sqrt{g_{00} (x)} \right)^{-1} \left( \int d \chi + L \right) \\
\widetilde \chi &= \beta \Gamma \sqrt{g_{00} (x)} \ c \left( \int d \tau + \tau_0 \right) + \ \ \Gamma \sqrt{g_{00} (x)} \left( \int d \chi + L \right)
\end{split}
\end{equation}
Since $\int d \chi = 0$ in the particle frame $S$, we set the following boundary conditions
\begin{equation}
\label{bc2} \int d \tau = 0 \quad \Rightarrow \quad \widetilde \tau = \int d \widetilde \tau = 0 \qquad \Rightarrow \qquad c \tau_0 = - \beta \left( g_{00} (x) \right)^{-1} L.
\end{equation}
and using (\ref{bc2}), the 2nd equation of (\ref{lorintp}) gives us:
$$\widetilde \chi = \beta \Gamma \sqrt{g_{00} (x)} \ c \int d \tau \ + \ \left( \sqrt{g_{00} (x)} \right)^{-1} \left( g_{00} (x) - \beta^2 \right) \Gamma L,$$
Now since the particle is in motion in $\widetilde S$, it moves from its starting position $\widetilde \chi_0 = \widetilde L$ 
with the velocity of frame $S$ wrt $\widetilde S$. Thus, using (\ref{tdilp}), the contracted length $\widetilde L$ is:
$$\widetilde \chi = \int d \widetilde \chi + \widetilde \chi_0 \ = \ \beta c \int d \widetilde \tau + \widetilde L \quad = \quad \beta c \Gamma \sqrt{g_{00} (x)} \int d \tau + \widetilde L,$$
$$\Rightarrow \quad \widetilde L = \widetilde \chi - \beta c \int d \widetilde \tau = \left( \sqrt{g_{00} (x)} \right)^{-1} \left( g_{00} (x) - \beta^2 \right) \Gamma L = \left( \sqrt{g_{00} (x)} \right)^{-1} \Gamma^{-1} L,$$
\begin{equation}
\label{lcontp} \boxed{\widetilde L = \left( \sqrt{g_{00} (x)} \right)^{-1} \Gamma^{-1} L.}
\end{equation}
For flat space, we set $g_{00} = 1$, and $\Gamma \longrightarrow \gamma$, which should restore the original 
time-dilation and length-contraction rules of special relativity for a free particle. \smallskip

Due to the lowered speed limit in comparison to the speed of light for special 
relativity, time intervals shall dilate and length intervals shall contract further in 
presence of a gravitational potential field.

\subsection{Gravitational redshift}

An alternative way to arrive at the time-dilation formula (\ref{tdilp}) is to write the 
metric in its two equivalent forms in the two frames $\widetilde S$ and $S$:
$$ds^2 = c^2 d \widetilde \tau^2 \left[ 1 + 2 \frac L{L_0} \right] = g_{00} \ c^2 d \tau^2 \qquad \Rightarrow \qquad d \widetilde \tau = \Gamma \sqrt{g_{00}} d \tau.$$
Thus, we can see from (\ref{repara}) that time dilation will occur under circumstances of 
either motion, or presence in a potential field, or due to both. Comparison to an equivalent metric in a frame in flat space without motion, we have 
$$\delta {\widetilde t} = \delta t \sqrt{1 + 2 \frac L{L_0} }.$$
If we consider free particle motion, we can see that:
\begin{equation}
\label{motdil} U({\bm x}) = 0 \qquad \Rightarrow \qquad \delta {\widetilde t} = \delta t \sqrt{1 - \left( \frac{|\bm v|}c \right)^2 }.
\end{equation}
On the other hand, for stationary observation, time dilation is caused by 
potential fields:
\begin{equation}
\label{potdil} {\bm v} = 0 \qquad \Rightarrow \qquad \delta {\widetilde t} = \delta t \sqrt{1 + \frac{2 U(\bm x)}{mc^2}}.
\end{equation}
This is confirmed by the theories of gravitational redshift that occurs as monochromatic light 
of a certain frequency in free space enters a gravitational field. If the time period of the light 
frequency in presence of a gravitational field is given by $T$, then according to (\ref{potdil}) 
$$U = - \frac{GMm}r \qquad \qquad U_{r = \infty} = 0,$$
$$T \equiv \delta t \qquad \Rightarrow \qquad T_0 = T_{(U = 0)} \equiv \delta {\widetilde t},$$
\begin{equation}
\label{gravrs1} T_0 = T \sqrt{1 - \frac{2 GM}{c^2 r}} \qquad \Rightarrow \qquad T = \frac{T_0}{\sqrt{1 - \frac{2 GM}{c^2 r}}}.
\end{equation}
If we define the Event Horizon radius as $r_0 = \frac{GM}{c^2}$, then the new frequency in the presence of 
a gravitational field according to (\ref{gravrs1}) is given by:
$$\nu = \frac1T = \frac1{T_0} \sqrt{1 - 2 \frac{r_0}r} \qquad \nu_\infty = \frac1{T_0},$$
\begin{equation}
\label{gravrs2} \nu = \nu_\infty \sqrt{1 - 2 \frac{r_0}r} \qquad \Rightarrow \qquad \Delta \nu = \nu_\infty \left( \sqrt{1 - 2 \frac{r_0}r} - 1 \right).
\end{equation}
Under a weak potential limit $2U << mc^2$ or $r_0 << r$, we will have:
$$\nu \approx \nu_\infty \left( 1 + \frac{U(\bm x)}{mc^2} \right) = \nu_\infty \left( 1 - \frac{r_0}r \right) \qquad \Rightarrow \qquad \Delta \nu \approx \frac{U(\bm x)}{mc^2} \nu_\infty = - \frac{r_0}r \nu_\infty.$$
Furthermore, we will also have:
\begin{equation}
\label{gravrs3} \frac{\nu_{r1}}{\nu_{r2}} = \sqrt{\frac{1 - 2 \dfrac{r_0}{r_1}}{1 - 2 \dfrac{r_0}{r_2}}}.
\end{equation}
which is confirmed in any literature on the topic of general relativity \cite{smcarroll}.
Next, we shall briefly summarize the formulation of relativistic Hamiltonian mechanics on 
curved spaces.

\section{Approximations and Limits}

When dealing with a relativistically described system, we often are required to apply 
approximations and limits to comform with the established non-relativistic formulation or the 
conventional relativistic formulation which uses $\gamma$ instead of $\Gamma$. This will confirm if we are 
on the right track in our analysis. Special relativity was formulated entirely for free particles. 
Now we shall explore various ways of applying a weak potential.

\numberwithin{equation}{subsection}

\subsection{Weak potential limit}

Under the weak potential approximation, we shall consider the case where
\begin{equation}
\label{wpappr} U (\bm{x}) << mc^2.
\end{equation}
Directly from (\ref{eom}), we can write that
$$\frac{d^2 \bm{x}}{d\widetilde{t}^2} = - \frac1m \left( \frac{d t}{d \widetilde{t}} \right)^2 \bm{\nabla} U (\bm{x}) \equiv - \frac{c^2}2 \left( \frac{d t}{d \widetilde{t}} \right)^2 \bm{\nabla} g_{00} (\bm{x}).$$
Applying weak potential approximation (\ref{wpappr}), we can say from (\ref{repara}) that
$$\frac{d \widetilde{t}}{d t} = \sqrt{1 + 2 \dfrac L{L_0}} \approx \sqrt{1 - \left( \dfrac vc \right)^2} = \gamma^{-1},$$
$$\therefore \qquad \frac{d^2 \bm{x}}{d\widetilde{t}^2} = - \frac{c^2}2 \bm{\nabla} \gamma^2 g_{00} (\bm{x}).$$
and since we are considering Lorentz transformations of space-time event intervals, and $g_{00} = 
g_{00} (\bm{x})$, we can say that
$$dt \longrightarrow d \widetilde{t} = \frac{d \widetilde{t}}{d t} dt \qquad \Rightarrow \qquad g_{00} \longrightarrow \widetilde{g}_{00} = \left( \frac{d \widetilde{t}}{d t} \right)^{-2} g_{00},$$
$$\therefore \qquad \widetilde{g_{00}} (\bm{x}) = \gamma^2 g_{00} (\bm{x}).$$
which lets us write the Lorentz-covariant equation of motion:
\begin{equation}
\label{ceom1} \frac{d^2 \bm{x}}{d\widetilde{t}^2} = - \frac{c^2}2 \bm{\nabla} \widetilde{g_{00}} (\bm{x}) \equiv - \frac1m \bm{\nabla} \widetilde{U} (\bm{x}).
\end{equation}
Now, we shall look at an alternate formulation with weak potentials.

\subsection{Semi-relativistic formulation}

Another way to describe the non-relativistic approximation $U (\bm{x}) << mc^2, |\bm v| << c$, of the 
metric (\ref{curlag}) using the expression of $\gamma$ from (\ref{repara}) is:
$$\mathcal{L} = - mc^2 \sqrt{\left(1 - \frac {|\bm v|^2}{c^2} \right) + \frac {2 U}{mc^2}} = - mc^2 \sqrt{\gamma^{-2} + \frac {2 U}{mc^2}} = - mc^2 \gamma^{-1} \sqrt{1 + \frac {2 U}{mc^2} \gamma^2}.$$
Binomially expanding the expression within the square-root gives us:
$$\mathcal{L} \approx - mc^2 \gamma^{-1} \left( 1 + \frac {U}{mc^2} \gamma^2 \right) = - mc^2 \gamma^{-1} - U \gamma,$$
\begin{equation}
\label{srlag1} \therefore \qquad \mathcal{L} \approx - mc^2 \gamma^{-1} - U \gamma.
\end{equation}
which is different from the form of the Lagrangian employed by Goldstein \cite{goldstein}. However, we 
must keep in mind that upon applying it into Euler-Lagrange equations, we get the semi-relativistic equation of motion:
\begin{equation}
\label{sreom1} m \frac{d \ }{d t} \left( \gamma {\bm v} \right) = - \gamma {\bm \nabla} U \qquad \xrightarrow{multiply \ \gamma} \qquad m \frac{d^2  {\bm x}}{d {\widetilde t}^2} = - \gamma^2 {\bm \nabla} U.
\end{equation}
which is the same equation of motion given by (\ref{ceom1}). Another approximation we can employ 
here is based on the comparison of the magnitude of factors paired with $\gamma^{-1}$ and $\gamma$ in (\ref{srlag1}) 
upon binomial expansion. In simple words, if $\gamma^{-1}$ and $\gamma$ in (\ref{srlag1}) are expanded to 1st order:
$$\gamma^{-1} \approx \left( 1 - \beta^2 \right)^{\frac12} \approx 1 - \frac{\beta^2}2, \qquad \gamma \approx \left( 1 - \beta^2 \right)^{- \frac12} \approx 1 + \frac{\beta^2}2, \qquad \text{where } \beta = \frac{|\bm v|}c.$$
we can see that higher order terms from expansion will become significant contributors depending on the factor 
multiplied to it. Now we can see that for weak potentials
$$mc^2 >> U (\bm x).$$
This means that at least 1st order contribution from $\gamma^{-1}$ will be significant in $mc^2 \gamma^{-1}$. On 
the other hand, the 1st order contribution from$\gamma$ will be insignificant in $U \gamma$. This analysis is 
elaborated as shown below:
\begin{equation}
\label{aprx} \therefore \quad \frac{|\bm v|}c = \beta << 1 \quad \Rightarrow \qquad 
\begin{cases}
mc^2 \gamma^{-1} \approx m \left( c^2 - \dfrac{|\bm v|^2}2 \right) \\ 
U \gamma \approx U + \dfrac U2 \beta^2
\end{cases} .
\end{equation}
Clearly, we can see that in (\ref{aprx}), the existence of the $\beta^2$ term in the potential energy part of 
(\ref{srlag1}) allows us to safely omit a part of of the Lagrangian for the limit $\beta << 1$. This means 
that we can say:
$$\lim_{\beta << 1} \frac U2 \beta^2 = 0 \qquad \Rightarrow \qquad U \gamma \approx U.$$
Thus, we can say that the semi-relativistic Lagrangian for low velocities can be written as:
\begin{equation}
\label{srlag2} \therefore \qquad \mathcal{L}_{sr} \approx - mc^2 \gamma^{-1} - U.
\end{equation}
Now this matches the form of the semi-relativistic Lagrangian (\ref{adhoc}) employed by Goldstein 
\cite{goldstein}. We can further proceed to say that if the same pattern of approximation is applied to the 
semi-relativistic equation of motion (\ref{sreom1}), we shall have:
$$m \frac{d^2  {\bm x}}{d {\widetilde t}^2} = - \gamma^2 {\bm \nabla} U \approx - \left( 1 + \beta^2 \right) {\bm \nabla} U.$$
Using the same approximation rule (\ref{aprx}), we can ignore the term with $\beta^2$ to write:
$$\lim_{\beta << 1} \beta^2 {\bm \nabla} U = 0 \qquad \Rightarrow \qquad \gamma^2 {\bm \nabla} U \approx {\bm \nabla} U,$$
\begin{equation}
\label{sreom2} \therefore \qquad \lim_{\beta << 1} m \frac{d^2  {\bm x}}{d {\widetilde t}^2} = - {\bm \nabla} U.
\end{equation}
This equation is thus nearly the same as the usual equation of motion known classically, except 
for the usage of proper time $\widetilde t$ in the particle frame instead of $t$. 
From this equation, we can write a conserved quantity given as:
\begin{equation}
\label{cnsvd} H = \frac m2 \bigg| \frac{d  \bm x}{d \widetilde t} \bigg|^2 + U.
\end{equation}
As stated, this formulation shall 
only apply in the low velocity limit for weak potentials. Now we shall look at the relativistic 
Hamiltonian formulation under weak potentials.

\subsection{Hamiltonian formulation under weak potential}

Under circumstances of a weak potential ($2 U (\bm{x}) << mc^2$) and low momentum $\left( \dfrac{|\bm p|}{mc} \approx 0 \right)$, another way to write the relativistic energy (\ref{relen1}) is:
$$\mathcal{E} = \sqrt{1 + \frac{2 U (\bm{x})}{mc^2}} \sqrt{|p|^2 c^2 + m^2 c^4} \approx \left(1 + \frac{U (\bm{x})}{mc^2} \right) \sqrt{|p|^2 c^2 + m^2 c^4},$$
$$\Rightarrow \qquad \mathcal{E} \approx \sqrt{|p|^2 c^2 + m^2 c^4} + U (\bm{x}) \sqrt{ \left( \frac{|p|}{mc} \right)^2 + 1} \approx \sqrt{|p|^2 c^2 + m^2 c^4} + U (\bm{x}),$$
\begin{equation}
\label{relen2} \therefore \qquad \mathcal{H} = \mathcal{E} \approx \sqrt{|p|^2 c^2 + m^2 c^4} + U (\bm{x}).
\end{equation}
Thus, the Hamilton's equations of motion (\ref{hameq1}) evolve into the form presented in \cite{babusci}:
\begin{equation}
\begin{split}
\dot{\bm x} &= \frac{\partial \mathcal{H}}{\partial \bm{p}} = \frac{c \bm{p}}{\sqrt{|p|^2 + m^2 c^2}}, \\
\dot{\bm p} &= - \frac{\partial \mathcal{H}}{\partial \bm{x}} = - \bm{\nabla} U (\bm{x}).
\end{split}
\end{equation}
Now we shall proceed to study the relativistic oscillator and its duality with the Kepler system.

\numberwithin{equation}{section}

\section{Bohlin-Arnold Duality}

Now that we have properly described the relativistic formulation for classical mechanical 
systems in a general scalar potential field, we shall now apply this formulation to two important 
mechanical systems frequently discussed in classical mechanics. They are the Hooke oscillator 
and Kepler systems. These systems are also dual to each other 
via a conformal transformation known as the Bohlin-Arnold-Vasiliev transformation.

\numberwithin{equation}{subsection}

\subsection{Relativistic 2D Isotropic Oscillator and Kepler systems}

One may ask how such a duality is a matter of concern here in the analysis of Newtonian gravity. While 
the Kepler potential is known to describe Newtonian gravity, it doesn't seem possible to find a Hooke's law 
potential that can be described as a result of curved space. The answer lies in the way Hooke's oscillator 
mechanics are applied in physics; around equilibrium points. In the study of planetary motion, one encounters 
equilibrium points known as Lagrange points that allow planets to maintain stable orbits. It is locally around 
these points that one will find single particles to exhibit Hooke oscillatory motion, whose potential function 
can be described as the curvature of the local spacetime. \\ \\
In a manner similar to (\ref{curmet}), the metric of the relativistic gravitational oscillator 
can be given by:
\begin{equation}
\label{metricosc} ds^2 = \left( 1 + \frac{k r^2}{mc^2} \right) c^2 dt^2 - dr^2 - r^2 \left( d \theta^2 + \sin^2 \theta \ d \varphi^2 \right).
\end{equation}
The Lagrangian corresponding to (\ref{metricosc}) according to (\ref{curlag}) would be
\begin{equation}
\label{rlagosc} \mathcal{L} = - mc^2 \sqrt{1 - \frac 2{mc^2} \left[ \frac{m \left[ \dot{r}^2 + r^2 \left( \dot{\theta}^2 + \sin^2 \theta \ \dot{\varphi}^2 \right) \right]}2 - \frac{kr^2}2 \right]}.
\end{equation}
For planar motion $\theta = \frac{\pi}2$, the momenta are given by:
\begin{equation} \label{mom1} 
\begin{split}
p_r &= mc \Gamma \dot{r} \qquad \qquad 
p_{\varphi} = mc \Gamma r^2 \dot{\varphi} = l, \\
\mathcal{E} &= p_r \dot{r} + p_{\varphi} \dot{\varphi} - \mathcal{L} = m c^2 \left( 1 + \dfrac{k r^2}{mc^2} \right) \Gamma, \\
\text{where } \qquad & \Gamma = \frac1{\sqrt{1 - \dfrac 2{mc^2} \left[ \dfrac{m \left( \dot{r}^2 + r^2 \dot{\varphi}^2 \right)}2 - \dfrac{kr^2}2 \right]}}.
\end{split}
\end{equation}
So the relativistic radial equation would be given by:
\begin{equation}
\label{intermed2osc} \frac{d \ }{d t} \left( \Gamma \dot{r} \right) = - \left( \frac{k r}{m} - r \dot{\varphi}^2 \right) \Gamma.
\end{equation}
while the angular equation is given by:
$$\dfrac{d p_\varphi}{d \tau} = \dfrac{d \ }{d \tau} \left( r^2 \dot{\varphi} \Gamma \right) = 0.$$
If we choose the proper time $\widetilde{t}$ as parameter
$$\Gamma = \dfrac{d \widetilde{t}}{d t} = \sqrt{1 - \dfrac 2{mc^2} \left[ \dfrac{m\left( \dot{r}^2 + r^2 \dot{\varphi}^2 \right)}2 - \dfrac{kr^2}2 \right]}.$$
then we can modify equation (\ref{intermed2osc}) in accordance with (\ref{eom}) to:
\begin{equation}
\label{eom1} m \left[ \dfrac{d^2 r}{d \widetilde{t}^2} - r \left( \dfrac{d \varphi}{d \widetilde{t}} \right)^2 \right] = - kr\left( \dfrac{d t}{d \widetilde{t}} \right)^2.
\end{equation}
For small oscillations, according to (\ref{srlag1}), the relativistic Lagrangian is:
\begin{equation}
\label{wlagosc} \mathcal{L} = - mc^2 \gamma^{-1} - \frac{k r^2}2 \gamma.
\end{equation}
The Euler-Lagrange equation of motion that we can derive from (\ref{wlagosc}) are:
\begin{equation}
\label{alteom1} \dfrac{d \ }{d t} \left( \gamma {\bm v} \right) = - \omega^2 \gamma \bm{x} \qquad \qquad \omega^2 = \frac km.
\end{equation}
which does not match the form presented in \cite{harvey}. However, for small oscillations during which the maximum velocities achieved are relatively small compared to the speed of light, we shall have according to (\ref{srlag2}) and (\ref{sreom2}):
\begin{equation}
\label{srosc}\mathcal{L}_{sr} \approx - mc^2 \gamma^{-1} - \frac{\omega^2}2 |\bm x|^2, 
\end{equation}
\begin{equation}
\label{alteom2} \dfrac{d \widetilde {\bm v}}{d \widetilde t} = - \omega^2 \bm x, \qquad \qquad \text{where } \omega^2 = \frac km.
\end{equation}
which matches the form presented in \cite{harvey}. \\

\noindent
Again, according to (\ref{curmet}), the relativistic Kepler system can be given by:
\begin{equation}
\label{metrickep} ds^2 = \left( 1 - \frac{2 G M}{r c^2} \right) c^2 dt^2 - dr^2 - r^2 \left( d \theta^2 + \sin^2 \theta \ d \varphi^2 \right).
\end{equation}
The Lagrangian corresponding to (\ref{metrickep}) according to (\ref{curlag}) would be
\begin{equation}
\label{rlagkep} \mathcal{L} = - mc^2 \sqrt{1 - \frac 2{mc^2} \left[ \frac{m \left[ \dot{r}^2 + r^2 \left( \dot{\theta}^2 + \sin^2 \theta \ \dot{\varphi}^2 \right) \right]}2 + \frac{G M m}r \right]}.
\end{equation}
For planar motion $\theta = \frac{\pi}2$, the momenta are given by:
\begin{equation} \label{mom2} 
\begin{split}
p_r &= mc \Gamma \dot{r} \qquad \qquad 
p_{\varphi} = mc \Gamma r^2 \dot{\varphi} = l, \\
\mathcal{E} &= p_r \dot{r} + p_{\varphi} \dot{\varphi} - \mathcal{L} = m c^2 \left( 1 - \dfrac{2 G M}{r c^2} \right) \Gamma, \\
\text{where } \qquad & \Gamma = \frac1{\sqrt{1 - \dfrac 2{mc^2} \left[ \dfrac{m \left( \dot{r}^2 + r^2 \dot{\varphi}^2 \right)}2 + \dfrac{G M m}r \right]}}.
\end{split}
\end{equation}
So the relativistic radial equation would be given by:
\begin{equation}
\label{intermed2kep} \frac{d \ }{d t} \left( \Gamma \dot{r} \right) = - \left( \frac{G M}{r^2} - r \dot{\varphi}^2 \right) \Gamma.
\end{equation}
while the angular equation is given by:
$$\dfrac{d p_\varphi}{d \tau} = \dfrac{d \ }{d \tau} \left( r^2 \dot{\varphi} \Gamma \right) = 0 \qquad \Rightarrow \qquad p_\varphi = r^2 \dot{\varphi} \Gamma = const.$$
showing that the angular momentum is conserved for radial forces. If we choose the proper 
time $\widetilde{t}$ as parameter
$$\Gamma = \dfrac{d \widetilde{t}}{d t} = \sqrt{1 - \dfrac 2{mc^2} \left[ \dfrac{m\left( \dot{r}^2 + r^2 \dot{\varphi}^2 \right)}2 + \dfrac{G M m}r \right]}.$$
then we can modify equation (\ref{intermed2kep}) in accordance with (\ref{eom}) to:
\begin{equation}
\label{eom2} m \left[ \dfrac{d^2 r}{d \widetilde{t}^2} - r \left( \dfrac{d \varphi}{d \widetilde{t}} \right)^2 \right] = - \frac{G M}{r^2} \left( \dfrac{d t}{d \widetilde{t}} \right)^2.
\end{equation}
For small oscillations, according to (\ref{srlag1}), the relativistic Lagrangian is:
\begin{equation}
\label{wlagkep} \mathcal{L} = - mc^2 \gamma^{-1} + \frac{G M m}r \gamma.
\end{equation}
The Euler-Lagrange equation of motion that we can derive from (\ref{wlagkep}) are:
\begin{equation}
\label{alteom3} \dfrac{d \ }{d t} \left( \gamma {\bm v} \right) = - \frac{G M m}{r^3} \gamma \bm{x}.
\end{equation}
Again, for relatively low velocities, we shall have according to (\ref{sreom2}):
\begin{equation}
\label{srkep}\mathcal{L}_{sr} \approx - mc^2 \gamma^{-1} + \frac{G M m}r, 
\end{equation}
\begin{equation}
\label{alteom4} \dfrac{d \widetilde {\bm v}}{d \widetilde t} = - \frac{G M m}{r^3} \bm x.
\end{equation}
We shall now briefly turn our attention to the Bohlin-Arnold duality to demonstrate semi-
relativistic Kepler-Hooke duality.

\subsection{Semi-Relativistic Kepler-Hooke duality}

The Kepler-Hooke duality established by Bohlin, Arnold and Vasiliev \cite{dual, bohlin} is a connection 
between the two mechanical systems which according to Bertrand's Theorem are the only 
two that are possible with closed, periodic orbits. This duality is established for the classical 
cases, but are not possible for the relativistic versions of the mechanical systems due to the 
$\Gamma$ factors involved in the equations of motion. Here, we show that it is again possible for the 
semi-relativistic equations of motion given by (\ref{alteom2}) and (\ref{alteom4}). \\ \\
The Bohlin transformation is a conformal map given by:
\begin{equation}
\label{conf} f: z \longrightarrow \xi = \left( z \right)^2 = R e^{i \phi} \qquad \Rightarrow \qquad z = \xi^{\frac 1 2}.
\end{equation}
Now we must note that another Noether invariant, the angular momentum will change form under this transformation. We re-parametrize to preserve the form of angular momentum.
$$l = r^2 \dot{\theta} = \vert z \vert ^2 \dot{\theta} = \vert \xi \vert^2 \phi' \qquad \Rightarrow \qquad \vert \xi \vert \frac{d \widetilde{\tau}}{d \widetilde t} \theta' = \vert \xi \vert^2 \theta',$$
\begin{equation}
\label{repar} \therefore \quad \widetilde t \longrightarrow \widetilde \tau : \frac{d \widetilde \tau}{d \widetilde t} = \vert \xi \vert.
\end{equation}
The velocity and acceleration transformation can be given using (\ref{conf}) and (\ref{repar}):
\[ \begin{split}
\dot{z} &= \frac 1 2 \frac{\vert \xi \vert}{\left( \xi \right)^{\frac 1 2}} {\xi}' = \frac 1 2 \left( \bar{\xi} \right)^{\frac 1 2} {\xi}', \\
\ddot{z} &= \frac 1 2 \vert \xi \vert \frac d {d \tilde{\tau}} \left\{ \left( \bar{\xi} \right)^{\frac 1 2} {\xi}' \right\} = \frac 1 2 \frac{\vert \xi \vert^2}{\left( \xi \right)^{\frac 1 2}} {\xi}'' + \frac 1 4 \left( \xi \right)^{\frac 1 2} \vert {\xi}' \vert^2.
\end{split} \]
Thus, the semi-relativistic equation of motion for oscillators (\ref{alteom2}) eventually becomes:
$$m \left\{ \frac 1 2 \frac{\vert \xi \vert^2}{\left( \xi \right)^{\frac 1 2}} {\xi}'' + \frac 1 4 \left( \xi \right)^{\frac 1 2} \vert {\xi}' \vert^2 \right\} = - k \left(\xi \right)^{\frac 1 2},$$
\begin{equation}
\label{transeom} \Rightarrow \qquad {\xi}'' = - \left( \frac 1 2 \vert {\xi}' \vert^2 + \frac {2k} m \right) \frac{\xi}{\vert \xi \vert^2}.
\end{equation}
Using the conserved quantity $H$ from (\ref{cnsvd}) for the oscillator system
$$H = \frac m2 | \dot z |^2 + \frac k2 | z |^2 = \frac m 4 \left( \frac{\left| {\xi'} \right|^2}2  + \frac {2k} m \right) \vert \xi \vert.$$
we can complete the transformation (\ref{transeom}) using $\left( \frac {\vert \xi' \vert^2 } 2 + \frac {2k} m \right) = \frac{4 H} m \frac 1 {\vert \xi \vert} = \kappa \frac 1 {\vert \xi \vert}$:
\begin{equation}
\therefore \qquad {\xi}'' = - \left( \frac {\vert {\xi}' \vert^2 } 2 + \frac {2k} m \right) \frac {\xi}{\vert \xi \vert^2} \equiv - \kappa \frac {\xi}{\vert \xi \vert^3}.
\end{equation}
showing that the transformation restores the central force nature of the system and produces 
the complex version of the Kepler equation (\ref{alteom4}). 

\numberwithin{equation}{section}

\section{Relativistic Lienard-type oscillator}

In the study of dynamical systems, the Lienard system is a 2nd order differential equation 
named after French  physicist Alfred-Marie Li\'enard \cite{polzai}. It is a very generalized way to describe 
1-dimensional motion under the influence of scalar potential and damping effects. Such differential 
equations were used to model oscillating circuits for applications in radios and vaccum 
tubes. For oscillatory systems, Li\'enard's Theorem, under certain assumptions assures uniqueness 
and existence of a limit cycle for the system. \\ \\
The equation of a damped 1-dimensional relativistic harmonic oscillator is:
$$\gamma^3 \ddot{x} + \alpha \gamma \dot{x} + \omega^2 x = 0 \qquad \qquad \gamma = \left( 1 - \frac{\dot{x}^2}{c^2} \right)^{-\frac12}.$$
In contrast, the damped relativistic Lienard equation is:
\begin{equation}
\label{lienard} \gamma^3 \ddot{x} + \gamma f(x) \dot{x} + g(x) = 0 \qquad \qquad \gamma = \left( 1 - \frac{\dot{x}^2}{c^2} \right)^{-\frac12}.
\end{equation}
Under reparametrization $t \longrightarrow d \widetilde{t} = \gamma^{-1} d t$, we have $x' = \frac{d x}{d \widetilde t} = \gamma \frac{d x}{d t} = \gamma \dot{x}$, letting us write:
$$\frac{d \ }{dt} \left( \gamma \dot{x} \right) = \gamma \ddot{x} + \gamma^3 \frac{{\dot x}^2}{c^2} \ddot{x} = \gamma^3 \ddot{x} \left( 1 - \frac{\dot{x}^2}{c^2} + \frac{{\dot x}^2}{c^2} \right) = \gamma^3 \ddot{x},$$
$$\therefore \qquad \gamma^3 \ddot{x} = \frac{d \ }{dt} \left( \gamma \dot{x} \right) = \frac{d \ }{dt} x' = \gamma^{-1} \frac{d \ }{d \widetilde t} x' = \gamma^{-1} x''.$$
Thus, under reparametrization, (\ref{lienard}) becomes
\begin{equation}
\label{step1} x'' + \gamma \left[ f(x) x' + g(x) \right] = 0.
\end{equation}
Remember that
$$\gamma^{-2} = 1 - \left( \frac{\dot x}c \right)^2 = 1 - \gamma^{-2} \left( \frac{x'}c \right)^2 \qquad \Rightarrow \qquad \gamma = \sqrt{1 + \left( \frac{x'}c \right)^2}.$$
Thus, when $\dfrac{x'}c \longrightarrow 0$, we will have $\gamma \longrightarrow 1$, which is the same as when $\dfrac{\dot x}c \longrightarrow 0$. 

\numberwithin{equation}{subsection}

\subsection{Integrability and Relativistic Chiellini  condition}

Now, the relativistic Chiellini condition is given by:
\begin{equation}
\label{chellini} \frac{d \ }{d x} \left( \frac gf \right) = - \alpha \left( 1 + \alpha \right) \gamma f(x).
\end{equation}
Thus, using (\ref{chellini}), and using the integrating factor $\Omega = \int^{\widetilde t} d \tau \ \gamma f(x)$, we can rewrite (\ref{step1}) as:
\begin{equation}
\label{step2} x'' + \gamma f(x) x' - \frac1{\alpha \left( 1 + \alpha \right)} \left( \frac gf \right) \frac{d \ }{d x} \left( \frac gf \right) = 0.
\end{equation}
$$\Rightarrow \quad 2 \text{e}^{\Omega} \left[ x' x'' + \gamma f(x) \left( x' \right)^2 \right] - \frac{2 \text{e}^{\Omega}}{\alpha \left( 1 + \alpha \right)}  \left( \frac gf \right) \left\{ \frac{d \ }{d x} \left( \frac gf \right) x' \right\} = 0,$$
\[ \begin{split}
\Rightarrow \quad \frac{d \ }{d \widetilde t} \left[ \left( \text{e}^{\Omega} ( x' )^2 \right) - \frac{\text{e}^{\Omega}}{\alpha \left( 1 + \alpha \right)} \frac gf x' - \frac{\text{e}^{\Omega}}{\alpha \left( 1 + \alpha \right)} \left( \frac gf \right)^2 \right] & \\ 
& \hspace{-4cm} + \frac1{\alpha \left( 1 + \alpha \right)} \left( \frac gf \right) \left[ \frac{d \ }{d \widetilde t} \left( \text{e}^{\Omega} x' \right) + \left( \frac gf \right) \frac{d \text{e}^\Omega}{d \widetilde t} \right] = 0.
\end{split} \]
Now, using the Chiellini condition (\ref{chellini}), and referring to (\ref{step1}) we can see that
\[ \begin{split}
\frac{d \ }{d \widetilde t} \left( \text{e}^{\Omega} x' \right) + \left( \frac gf \right) \frac{d \text{e}^{\Omega}}{d \widetilde t} &= \text{e}^{\Omega} \left[ x'' + \left\{ x' + \left( \frac gf \right) \right\} \left( \gamma f(x) \right) \right], \\
&= \text{e}^{\Omega} \left[ x'' + \gamma \left( f(x) x' + g(x) \right) \right] = 0,
\end{split} \]
$$\therefore \qquad \frac{d \ }{d \widetilde t} \left[ \text{e}^{\Omega} \left\{ ( x' )^2 - \frac1{\alpha \left( 1 + \alpha \right)} \frac gf \left( x' + \frac gf \right) \right\} \right] = 0.$$
Thus, we have a conserved quantity given by
\begin{equation}
\label{fint} I = \text{e}^{\Omega} \left[ \gamma^2 ( \dot x )^2 - \frac1{\alpha \left( 1 + \alpha \right)} \frac gf \left( \gamma \dot x + \frac gf \right) \right], \qquad \qquad \Omega = \int^{\widetilde t} d \tau \ \gamma f(x).
\end{equation}
Thus, we have a conserved quantity for a relativistic Li\'enard system. To solve it for the damped oscillator, we write $f(x) = \kappa$, $g(x) = x$.

\subsection{Metric and Lagrangian}

To deduce the metric from the relativistic equation of motion for a damped system (\ref{lienard}), 
we shall execute a more elaborate 5-step procedure than the 3-step procedure for undamped 
systems, given by:

\begin{enumerate}
\item Convert the equation (\ref{lienard}) to the non-relativistic version (with $\gamma = 1$).
\begin{equation}
\label{dmp1} \ddot{\bm x} + \alpha \dot{\bm x} + \frac1m {\bm \nabla} U = 0.
\end{equation}
\item Determine the reparametrization factor $\text{e}^{\Omega}$ by converting (\ref{dmp1}) to the form (\ref{classeom}).
$$\frac{d \ }{d t} \left( \text{e}^{\Omega} \dot{\bm x} \right) + \frac{\text{e}^{\Omega}}m \nabla U = 0, \qquad \qquad \Omega = \int^t d t' \ \alpha,$$
\begin{equation}
\label{dmp2} \Rightarrow \qquad {\bm x}'' = - \frac{\text{e}^{2 \Omega}}m {\bm \nabla} U = - \frac{c^2 \text{e}^{2 \Omega}}2 {\bm \nabla} g_{00}, \qquad \qquad {\bm x}' = \text{e}^{\Omega} \dot{\bm x}, \qquad d \tau = \text{e}^{- \Omega} dt.
\end{equation}
\item Deduce the undamped potential $U$ from (\ref{dmp2}) by factoring out $\text{e}^{2 \Omega}$.
\item Comparing (\ref{dmp2}) to (\ref{effeom}), we can formulate the damped effective Lagrangian $L_d$ for reparameterized time $\tau$, just as (\ref{efflag}) can be derived from (\ref{effeom}).
$${\bm x}'' = - \frac{c^2 \text{e}^{2 \Omega}}2 {\bm \nabla} g_{00} \quad \longrightarrow \quad L_d = \frac m2 \left[ \big| {\bm x}' \big|^2 - \text{e}^{2 \Omega} c^2 g_{00} (\bm{x}) \right],$$
\begin{equation}
\label{dmp3} L_d = \frac m2 \text{e}^{2 \Omega} \left[ \big| \dot{\bm x} \big|^2 - c^2 g_{00} (\bm{x}) \right].
\end{equation}
\item Deduce the classical damped effective Lagrangian $L_{ed}$ from (\ref{dmp3}) by multiplying $\text{e}^{- \Omega}$, and write the damped metric $ds^2_d$ from it.
\[ \begin{split} 
\text{relativistic}: &\qquad d S = d \tau \ {\mathcal L} = \text{e}^{- \Omega} d t \ {\mathcal L} \\
\text{effective classical}: &\qquad d S_{eff} = d \tau \ L_d = \text{e}^{- \Omega} d t \ L_d = dt \ L_{ed}
\end{split} \]
$$L_d = - \frac m2 \left( \frac{ds}{d \tau} \right)^2 = - \frac m2 \text{e}^{2 \Omega} \left( \frac{ds}{dt} \right)^2,$$
$$L_{ed} = \text{e}^{- \Omega} L_d = - \frac m2 \text{e}^{- \Omega} \left( \frac{ds}{d \tau} \right)^2 = - \frac m2 \left( \frac{ds_d}{d t} \right)^2 = \frac m2 \text{e}^{\Omega} \left[ \big| \dot{\bm x} \big|^2 - c^2 g_{00} (\bm{x}) \right],$$
\begin{equation}
\label{dmpmet} ds^2_d = \text{e}^{\Omega} \left[ c^2 g_{00} (\bm{x}) dt^2 - \left| d {\bm x} \right|^2 \right].
\end{equation}
\end{enumerate}
Looking back at (\ref{lienard}), we can take the non-relativistic version of the equation and say that
\[ \begin{split} 
\ddot{x} &+ f(x) \dot{x} + g(x) = 0, \\
\therefore \qquad \frac{d \ }{d t} \left( \text{e}^{\Omega} {\dot x} \right) &= - \text{e}^{\Omega} g(x) \qquad \qquad \Omega = \int^t d \widetilde{t} \ f(x).
\end{split} \]
If we reparametrize as $d \tau = \text{e}^{-\Omega} d t$ then we will have using the non-relativistic version of 
Chiellini condition (\ref{chellini})
$$\frac{d \ }{d x} \left( \frac gf \right) = - \alpha \left( 1 + \alpha \right) f(x),$$
$$\frac{d^2 x}{d \tau^2} = - \text{e}^{2 \Omega} f(x) \frac gf = \frac{\text{e}^{2 \Omega}}{2 \alpha \left( 1 + \alpha \right)} \frac{d \ }{d x} \left[ \left( \frac gf \right)^2 \right] = - \frac{c^2}2 \frac{d \ }{d x} \left[ \text{e}^{2 \Omega} \left\{ 1 - \frac1{c^2 \alpha \left( 1 + \alpha \right)} \left( \frac gf \right)^2 \right\} \right],$$
\begin{equation}
\label{repardmp} \therefore \qquad \frac{d^2 x}{d \tau^2} = - \frac{c^2}2 \frac{d \ }{d x} \left[ \text{e}^{2 \Omega} \left\{ 1 - \frac1{c^2 \alpha \left( 1 + \alpha \right)} \left( \frac gf \right)^2 \right\} \right].
\end{equation}
Arguing that $x = x(t)$ and from (\ref{repardmp}) we have the undamped potential $U ({\bm x})$ and $g_{00} ({\bm x})$: 
\begin{equation}
\label{pot} U ({\bm x}) = - \frac m{2 \alpha \left( 1 + \alpha \right)} \left( \frac gf \right)^2 \qquad \Rightarrow \qquad g_{00} ({\bm x}) = \left[ 1 - \frac 1{c^2 \alpha \left( 1 + \alpha \right)} \left( \frac gf \right)^2 \right].
\end{equation}
Using (\ref{pot}), the damped effective Lagrangian according to (\ref{dmp3}), is given from (\ref{repardmp}) as:
$$L_d = \frac{m \text{e}^{2 \Omega}}2 \left[ \left( \frac{d x}{d t} \right)^2 - c^2 \left\{ 1 - \frac1{c^2 \alpha \left( 1 + \alpha \right)} \left( \frac gf \right)^2 \right\} \right].$$
Now, we can rewrite this as
\begin{equation}
\label{led} \Rightarrow \qquad L_{ed} = \text{e}^{- \Omega} L_d = \frac m2 \text{e}^{\Omega} \left[ \left( \frac{d x}{d t} \right)^2  - c^2 \left\{ 1 - \frac1{c^2 \alpha \left( 1 + \alpha \right)} \left( \frac gf \right)^2 \right\} \right].
\end{equation}
and finally, the metric with drag factor from (\ref{led}) is (\ref{dmpmet}) given by:
\begin{equation}
\label{lienmet} \left( \frac{d s}{d t} \right)^2 = - \frac2m L_{ed} \qquad \Rightarrow \qquad ds^2 = \text{e}^{\Omega} \left\{ 1 - \frac1{c^2 \alpha \left( 1 + \alpha \right)} \left( \frac gf \right)^2 \right\} c^2 d t^2 - \text{e}^{\Omega} dx^2.
\end{equation}
where we can restore the integrating factor $\Omega$ back to the relativistic version $\Omega = \int^t d \widetilde t \ \gamma f(x)$. \\ \\
This further goes to show that damped mechanical systems can be described by a metric that 
is spatially isotropic with a non-unity co-efficient. Furthermore, if it is a solution of Einstein's 
equations, then we should have:
$$\text{e}^{- 2 \Omega} = 1 - \frac1{c^2 \alpha \left( 1 + \alpha \right)} \left( \frac gf \right)^2.$$
so from (\ref{lienmet}) we get the metric:
$$ds^2 = \frac{c^2 d t^2}{\text{e}^{\Omega}} - \text{e}^{\Omega} dx^2.$$
This topic will be elaborated upon in greater detail in another project where such spaces are 
studied as damped mechanical systems. \footnote{Contact Hamiltonian mechanics \cite{bct} extend symplectic Hamiltonian mechanics \cite{arnold}, geometrically describing non-dissipative and dissipative systems, eg.: thermodynamics \cite{blmn}, mesoscopic dissipative mechanical systems \cite{grmela}, and mechanical systems drawing energy from a reservoir. 

\smallskip

\noindent
The Hamiltonian is effectively provided by $I$ in (\ref{fint}). If we define a new variable $s$, as in \cite{carguha}, ignoring the decay-countering factor $\text{e}^\Omega$ gives the decaying Hamiltonian embedded in (\ref{fint}):
$$s := - \frac1{\alpha \left( \alpha + 1 \right)} \left( \frac gf \right) \gamma {\dot x}, \qquad H =  (\gamma \dot x )^2 - \frac1{\alpha \left( 1 + \alpha \right)} \left( \frac gf \right)^2 + s.$$
From the relativistic non-decaying Lagrangian of (\ref{lienmet}), the relativistic momentum under weak potential approximates to $p \approx \gamma \dot x$, which we can replace in $s$ and $H$ defined above to rewrite them, from which we can see that
$$\dot s = f(x) \left[ \frac{\partial H}{\partial p} p - H \right].$$
However, as stated in \cite{carguha}, unless $f(x) = const$, we cannot recast the Li\'enard equation in 
contact form, which is simply the damped oscillator.}

\section{Conclusion}

The form of relativistic Lagrangian and equations of motion in \cite{harvey} also used by Goldstein 
\cite{goldstein}, do not match with the formulations derived directly from solving for the geodesic from 
the spacetime metric. The latter formulation clearly reproduces the time dilation effects in a 
potential field as observed in the phenomena of gravitational redshifts \cite{smcarroll}, while the former 
does not. The Hamiltonian formulation shown in \cite{babusci} applies only for weak potentials and low 
momentum. We effectively managed to write a Lorentz-covariant form of equations of motion 
in (\ref{coveom}) for geodesics that does not match the classic non-relativistic form of such equations 
and describe a relativistic deformation of the Euler-Lagrange equation. 

Since the familiar Lorentz transformation was designed for the special case of special 
relativity, where we deal with free particles without any potentials, it will not suffice in a more 
general case where a particle accelerates under the influence of potentials. The solution was to 
formulate a modified Lorentz transformation that will locally leave the metric invariant. The 
fact that such a transformation is limited to work locally is not surprising, given that 
according to the Equivalence principle, a curved space is locally diffeomorphically equivalent 
to a flat space, where a regular Lorentz transformation would be valid.

The Bohlin-Arnold duality \cite{dual, bohlin} is not valid for the relativistic oscillator under this direct 
formulation due to the $\Gamma$ factors in the equation of motion. However, it is possible to use 
a semi-relativistic approximation for weak potential and low velocity to produce the type of 
semi-relativistic Lagrangian in \cite{harvey, goldstein}. Using such approximation, and replacing the time with 
the proper time in the particle frame, the semi-relativistic equation of motion takes the classical 
form, allowing the Bohlin-Arnold duality to be valid under this approximation.

We have written the relativistic version of the Li\'enard equation, and deduced its conserved 
quantity, Lagrangian and metric. The conserved quantity was derived after first redefining the 
Chiellini integrability condition relativistically. The metric derived for the Li\'enard oscillator 
implies metrics that are spatially isotropic with non-unity co-efficient function. Thus, if the 
Schwarzschild, and most solutions to Einstein's equations are rewritten to spatially isotropic 
co-ordinates, they shall be found to exhibit damped motion. This implies that the dynamics 
of such solutions can be examined as some form of the relativistic Li\'enard mechanical system.

So far, this article helps to revise and generalize our fundamental formulation of relativity, 
while showing how familiar results can still be reproduced under suitable approximations. 
The results of this article can be applied in future along the direction of canonical ADM gravity 
and relativistic quantum mechanics as covered in articles like \cite{lusanna, acl} respectively.

\section*{Acknowledgement}

We thank Professors \'Angel Ballesteros and Alexei A. Deriglazov for their 
correspondence and advice, which were invauable in developing this article, and Prof. 
Gary Gibbons furthermore for his collaboration in \cite{cgg, cgg1} and encouragement. 
Furthermore, we also wish to express special gratitude to the anonymous referee 
whose valuable advice and questions helped improve and expand the content of 
this manuscript. The research of Partha Guha is supported by FAPESP through 
Instituto de Fisica de S\~ao Carlos, Universidade de Sao Paulo with grant number 
2016/06560-6.

\end{document}